\documentclass[conference]{IEEEtran}
\usepackage{fancyhdr}
\IEEEoverridecommandlockouts

\usepackage[numbers,sort&compress]{natbib} 
\usepackage{amsmath,amssymb,amsfonts}
\usepackage{graphicx}
\usepackage{booktabs}
\usepackage{multirow}
\usepackage{xcolor}
\usepackage{nicefrac}
\usepackage{microtype}
\usepackage{textcomp}
\usepackage{url}
\usepackage{hyperref}
\usepackage{placeins}
\usepackage{tikz}
\usetikzlibrary{arrows.meta,positioning,shapes.geometric}
\usepackage{enumitem}
\usepackage{xspace}
\usepackage{booktabs}
\usepackage{graphicx}
\usepackage{tikz}
\usepackage{fontawesome5}
\newcommand{\circled}[1]{\tikz[baseline=(char.base)]{
    \node[shape=circle,draw,inner sep=0.5pt,minimum size=2mm] (char) {\tiny #1};}}
\newcommand{\MainAgent}{\textsc{EnvAgent}\xspace}
\newcommand{\EnvBench}{\textsc{EnvBench}\xspace}

\begin{document}

\title{EnvAgent: Multi-Agent Approach to Building Science Environments in the Clouds}
\author{\IEEEauthorblockN{Yiming Cheng}
\IEEEauthorblockA{\textit{The University of Chicago} \\
Chicago, USA \\
eaminchan@uchicago.edu}
\and
\IEEEauthorblockN{Chris Low}
\IEEEauthorblockA{\textit{The University of Chicago} \\
Chicago, USA \\
chrislowzhengxi@uchicago.edu}
\and
\IEEEauthorblockN{Binrui Huang}
\IEEEauthorblockA{\textit{The University of Chicago} \\
Chicago, USA \\
binruih@uchicago.edu}
\and
\IEEEauthorblockN{Mengshi Zhang}
\IEEEauthorblockA{\textit{TensorBlock, Inc.} \\
USA \\
mengshiz@tensorblock.co}
\and
\IEEEauthorblockN{Zihao Chen}
\IEEEauthorblockA{\textit{TensorBlock, Inc.} \\
USA \\
zihaoc@tensorblock.co}
\and
\IEEEauthorblockN{Kate Keahey}
\IEEEauthorblockA{\textit{Argonne National Laboratory /} \\
\textit{The University of Chicago} \\
Chicago, USA \\
keahey@mcs.anl.gov}
\and
\IEEEauthorblockN{Paul Marshall}
\IEEEauthorblockA{\textit{The University of Chicago} \\
Chicago, USA \\
marshalp@uchicago.edu}
\and
\IEEEauthorblockN{Kexin Pei}
\IEEEauthorblockA{\textit{The University of Chicago} \\
Chicago, USA \\
kpei@uchicago.edu}
}
\maketitle
\thispagestyle{fancy}
\lhead{}
\rhead{}
\chead{}
\lfoot{\footnotesize{
SC26 Workshops, November 15-20, 2026, Chicago, Illinois, USA
\newline 979-8-3195-1221-5/26/\$31.00 \copyright 2026 IEEE}}
\rfoot{}
\cfoot{}
\renewcommand{\headrulewidth}{0pt}
\renewcommand{\footrulewidth}{0pt}
\begin{abstract}
AI agents are rapidly transforming scientific computing and broader software-intensive workflows, but their ability to generate, test, and execute code rests on a more fundamental prerequisite: a correctly configured execution environment. 
Automating environment construction across repositories remains difficult because projects encode heterogeneous and often incomplete requirements for operating systems, language toolchains, dependencies, and build procedures. 
Existing workflows thus depend on developer-maintained Dockerfiles and setup scripts, making environment preparation labor-intensive and difficult to scale, and potentially biasing scientific evaluations.

We present \MainAgent, an agentic framework that automatically synthesizes, builds, and verifies repository-level Docker environments. 
Rather than relying exclusively on generic command-line interaction or narrowly specialized tools that encode fixed assumptions, \MainAgent combines deterministic utilities for Docker execution and state management with coordinated agents for repository analysis, planning, Dockerfile synthesis, and iterative failure diagnosis and repair. 
We also introduce \EnvBench, a benchmark of 54 scientific artifacts and real-world open-source repositories spanning seven programming languages drawn from Chameleon Trovi and Multi-SWE-bench. 
For each repository, we manually curate a declarative rubric specifying its required structure, configuration, and functionality without prescribing a particular implementation, and a shared evaluator applies 2,108 weighted checks consistently across generated environments. 
Across four matched language-model settings, \MainAgent constructs an average of 43.5 executable environments out of 54, compared with 23.5 for the Codex and Claude Code baselines, and achieves 62.02 versus 24.19 on a composite score combining executability and rubric satisfaction. 
These results show that environment-specific agent orchestration substantially improves automated environment construction. 
Together, our findings highlight the value of environment-specific orchestration and rubric-based evaluation for developing more reliable and scalable environment-construction agents.
\end{abstract}


\section{Introduction}

The emergence of large language model (LLM)-based agents has revolutionized software development practices, enabling automated code generation, bug fixing, and program synthesis at unprecedented scales ~\cite{liu2024marscode,chen2023teaching,jimenez2023swe, kang2025explainable,lin2025eco,chen2021evaluating, cheng2025codemenv, grishina2025fully, liang2024large, hou2024large}. 
These intelligent agents demonstrate remarkable capabilities in understanding and manipulating code at various levels of abstraction, providing substantial automation benefits across diverse computational software engineering tasks including code modification, performance optimization, and security analysis.
Yet turning these capabilities into reliable computational results rests on a more fundamental prerequisite: the target software must build and run in a correctly configured execution environment.
For scientific workflows, this environment is part of the experimental artifact because it determines whether results can be reproduced and whether execution-based feedback can be trusted~\cite{keahey_2025_15306610, guo2025deepseek}.


Despite this critical role, environment construction is commonly treated as an input to agentic workflows rather than an automation target.
Existing workflows and benchmarks often rely on repository-specific Dockerfiles and setup scripts maintained by developers or benchmark creators~\cite{jimenez2023swe,cheng2025codemenv}.
These artifacts may be incomplete or outdated and can require substantial manual effort to repair and maintain.
Moreover, many scientific artifacts and research prototypes are released as code repositories without complete, executable procedures for reconstructing their environments and reproducing the reported results.
When their environments cannot be reconstructed reliably, these artifacts become difficult to validate or reuse, weakening the executable basis for verifying the corresponding scientific results.
This limitation also narrows the pool of realistic artifacts available for developing and evaluating scientific AI agents, including agents that translate papers into implementations or attempt to reproduce reported experiments~\cite{liu2026lifescibench,seo2026paper2code,starace2025paperbench}.


Automating this missing step requires substantially more than installing a declared list of packages.
Environment requirements are distributed across README files, dependency manifests, build scripts, continuous-integration configurations, and source code, and these sources may be incomplete or mutually inconsistent.
For example, a project may declare its language-level packages but omit a required system library, while a prebuilt dependency may require a newer version of \texttt{glibc} than the selected base image provides.
Two libraries may also require incompatible versions of a shared native runtime, requiring a workaround that is documented by neither library.
Moreover, a Docker image that builds and starts is not necessarily correct.
Tests may be silently skipped because optional dependencies are missing, or a critical workflow may fail only when a particular execution path is reached.
Such latent configuration errors can create substantial debugging effort and, more seriously, compromise the reproducibility and validity of downstream results.


\vspace{.1cm}\noindent\textbf{Challenges for directly applying AI agents.}
The diversity of repository requirements creates a tension between generality and specialization when automating environment construction.
General-purpose code agents such as Codex and Claude Code preserve flexibility by letting LLMs directly operate through generic, command-line tools, e.g., \texttt{cat}, \texttt{mv}, etc.
However, they leave the LLMs responsible for discovering the environment-construction workflow, composing these low-level operations, and maintaining state across repeated build failures.
More sophisticated tools, e.g., template-based Dockerfile generators, can encode more of the workflow and alleviate the burden on LLMs to assemble the low-level tools, but are often rigid by assuming fixed build systems or repository structures.
These assumptions make them difficult to generalize to diverse projects with native dependencies and nonstandard setup procedures.
Our key observation is that environment construction combines a \emph{stable operational structure} with \emph{repository-specific decisions}.
Specifically, operations such as initializing a workspace, building and running containers, capturing logs, and tracking execution state can be implemented deterministically, whereas dependency inference, configuration planning, and failure diagnosis require contextual reasoning.


\vspace{.1cm}\noindent\textbf{Our approach.}
Based on this observation, we present \MainAgent, a multi-agent workflow for automatically constructing repository-level execution environments.
By delegating key operations to dedicated LLMs , \MainAgent combines the reliability of deterministic tools and the flexibility of LLMs for diverse scenarios.
Specifically, \MainAgent assigns stable environment-management operations to deterministic utilities and uses coordinated LLM agents to inspect repository artifacts, formulate a setup plan, synthesize a Dockerfile, and diagnose and repair failures.
After each construction attempt, \MainAgent captures build and runtime feedback, summarizes the current state, and uses this information to guide the next attempt.
This task-specific division of responsibilities structures the environment-construction process without hard-coding assumptions about a particular programming language, package manager, or build system.


\vspace{.1cm}\noindent\textbf{Our benchmark.}
Environment construction also presents a distinct evaluation challenge.
Build and launch success often only checks that the repository is executable after its environment is constructed, but not that it contains the configuration and functionality required by the repository.
Traditional software tests generally presuppose a working environment and do not systematically characterize which environment requirements have been satisfied.
We thus introduce \EnvBench, a benchmark containing 54 scientific artifacts and real-world open-source repositories drawn from Chameleon Trovi and Multi-SWE-bench and spanning seven programming languages.
For each repository, we construct a declarative rubric that specifies the required environment structure, configuration, and functionality without prescribing how these requirements must be implemented.
A shared evaluation engine applies 2,108 weighted checks expressed through reusable checker primitives.
While there are approaches leveraging LLMs as a judge for evaluation~\cite{gu2026survey}, we deliberately rely on manual process to curate these rubrics to ensure reliability.


We evaluate \MainAgent on \EnvBench using Chameleon Cloud, an OpenStack-based research testbed \cite{keahey2020lessons}.
Across the fixed model settings, \MainAgent constructs an average of 43.5 executable environments out of 54, whereas the agentic baselines like Codex and Claude Code construct 23.5.
\MainAgent also achieves 62.02, compared with 24.19 for the baselines, on a composite score combining environment executability and rubric satisfaction.
These gains require 2--10$\times$ greater token consumption, exposing a tradeoff between computational cost and the iterative diagnosis needed to configure difficult repositories reliably.
As a preliminary extension, we also explore whether requirements inferred from a repository can guide the provisioning of compatible cloud resources, one step before setting up the execution environment when the repositories need to be deployed on the cloud platform (Section~\ref{sec:eval}).
This work makes the following contributions:

\begin{itemize}[leftmargin=*]
\item We introduce \MainAgent, an agentic workflow that combines deterministic environment-management utilities with coordinated agents to synthesize, build, diagnose, and verify repository-level Docker environments to support scientific evaluation and reproducibility.

\item We develop \EnvBench, a human-curated benchmark of 54 repositories and 2,108 weighted checks for evaluating both environment executability and satisfaction of repository-specific functionality requirements.

\item We conduct various comparisons with frontier agent frameworks, such as Codex and Claude Code, using the same LLMs, and demonstrate promising preliminary improvements in environment construction, while sharing more lessons learned for future works on automated execution environment construction and configurations.
\end{itemize}

\section{Overview and Motivation}

\vspace{.1cm}\noindent\textbf{Problem formulation.}
We formulate the repository-level environment construction as follows.
Given a source repository and a target machine, construct a Dockerfile that provides the software stack required to build and execute the repository.
Our primary task begins after the target machine becomes available, although the constructed environment must remain compatible with its operating system, processor architecture, and other hardware resources.
We also include a preliminary study when these hardware resources need to be automatically configured, e.g., in a cloud environment (Section~\ref{subsec:moving-to-cloud}).
Cloud resource provisioning and multi-node orchestration are broader concerns that we consider only as preliminary extensions.
Importantly, a successful environment should not just build and start, but also provide the dependencies, tools, configuration, and functionality expected by the repository.

\vspace{.1cm}\noindent\textbf{Key challenges.}
The key challenge of environment construction is that its requirements rarely appear in a single well-defined authoritative specification.
Instead, the steps and requirements are often distributed across README files, dependency configurations, build scripts, continuous-integration configurations, container specifications, and even the artifacts in source code, e.g., developer comments, docstrings, etc.
These sources may also be incomplete or mutually inconsistent.
For example, a package manifest may declare language-level dependencies while omitting a required compiler or system library, and a prebuilt dependency may require a newer version of \texttt{glibc} than the selected base image provides.
A repository may also combine multiple ecosystems, e.g., a Python package that builds a C/C++ extension or invokes an external system tool not managed by the Python package manager.

These constraints are interdependent and often non-monotonic.
Installing or upgrading one package may resolve the current error while introducing a conflict with another dependency.
The order of configuration operations can also matter, e.g., a compiler or system header must be installed before a native dependency can be built.
As each construction attempt exposes only the failures reached under the current configuration, environment construction naturally requires an iterative process of execution, diagnosis, and revision~\cite{rahman2024impact,nazario2025mitigating}.
These challenges have made automated solutions extremely challenging that manual tuning has been the only way for this task.
Unfortunately, a manually constructed environment does not by itself provide a repeatable specification from which the experiment can be reproduced.
AI-based techniques, especially with recent advance in LLM agents, have thus become a promising direction given its capabilities from handling much larger space of uncertainties~\cite{keahey_2025_15306610}.


\vspace{.1cm}\noindent\textbf{Selective specialization for environment construction.}
Existing off-the-shelf agents provide incomplete solutions to this problem.
General-purpose code agents preserve flexibility by operating through generic command-line and file-editing tools, but leave the LLMs responsible for decomposing the construction workflow, composing low-level operations, interpreting build logs, and maintaining state across repeated failures.
On the other hand, highly specialized agent tools, e.g., analyzing source code comments, can reduce this reasoning burden, but often encode assumptions about particular repositories, package managers, or build systems.
Such assumptions make them difficult to apply to diverse repositories with mixed requirements, dependencies, or nonstandard setup procedures.

Environment construction thus benefits from selective rather than complete specialization.
Stable operations, e.g., workspace initialization, Docker execution, log collection, state tracking, and termination, can be implemented as deterministic utilities.
Repository-dependent decisions, e.g., inferring missing dependencies, selecting compatible versions, planning configuration steps, and repairing failures, require contextualized reasoning hard to be canonicalized.
This division reduces the burden of repeatedly assembling low-level operations without constraining the agent to a fixed repository structure or software ecosystem.

\vspace{.1cm}\noindent\textbf{Need for environment-level verification.}
Docker build and launch success are insufficient indicators of environment correctness.
A container may build successfully while containing the wrong dependency version, omitting a required executable or environment variable, or silently failing tests because an optional dependency is absent.
Traditional software testing also presuppose a usable environment and may not identify which environment requirements are satisfied or missing.
Evaluating environment construction thus requires explicit, repo-specific criteria covering structure, configuration, and functionality.
These criteria should specify what the environment must provide without prescribing how the agent constructs it, allowing different but functionally equivalent environments to be evaluated consistently.

These observations motivate our environment-construction-specific agentic workflow that combines deterministic tool support with flexible repo-level reasoning and iterative repair.
They also motivate a human-curated, rubric-based benchmark that evaluates both whether an environment executes and whether it satisfies the repository's actual requirements.
\MainAgent and \EnvBench instantiate these two requirements, repspectively.

\section{Method}

\begin{figure}[!t]
  \centering
  \includegraphics[width=\linewidth]{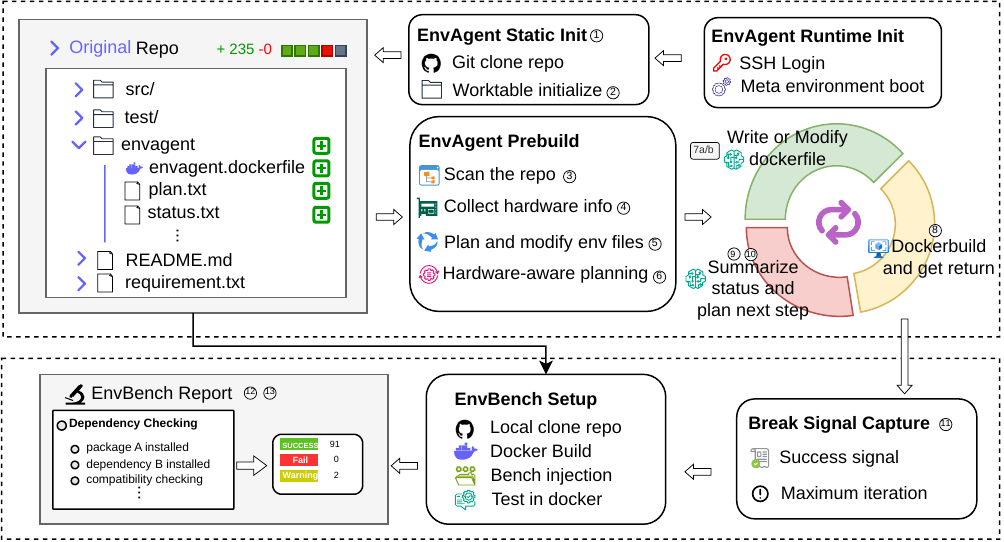}
\caption{\small Overview of \MainAgent on \EnvBench. 
Given a target repository, \MainAgent coordinates specialized agents to analyze the repository, plan the environment, construct and revise a Dockerfile, and interpret execution feedback. 
\EnvBench provides the repository for which \MainAgent is tasked to construct the environment, and declarative evaluation rubrics to compute fine-grained evaluation signals. 
The numbered workflow components are detailed in Table~\ref{tab:EnvAgent_agents}.}
\label{fig:overview}
\end{figure}

This section overviews the workflow of \MainAgent and its interaction with \textsc{EnvBench}. Figure~\ref{fig:overview} shows an
overview.

\subsection{Overview of \MainAgent}
\label{sec:EA}

\MainAgent orchestrates multiple specialized agents through a three-phase pipeline comprising \emph{preprocessing and planning}, \emph{execution and feedback}, and \emph{summarization and context management}. 
At each iteration, \MainAgent employs the workflow following ReAct~\cite{yao2022react}, where agents generate reasoning traces and execute actions, then incorporate environmental feedback to refine subsequent operations.
 
\vspace{.1cm}\noindent\textbf{Phase 1: Preprocessing and planning.} 
\MainAgent initiates environment configuration through coordinated context collection by the following specialized agents. 
The \emph{Scanning Agent} traverses the repository structure to construct a comprehensive file tree and identifies environment-relevant artifacts (e.g., configuration files, dependency manifests, build scripts). 
Concurrently, the \emph{Hardware Checking Agent} performs system introspection to gather hardware specifications including CPU architecture, GPU models and memory capacity, system RAM, disk space, and operating system details. 
Subsequently, the \emph{Planning Agent} synthesizes the collected repository context and hardware constraints to formulate a comprehensive environment configuration strategy, generating a step-by-step execution plan for containerized deployment.

\vspace{.1cm}\noindent\textbf{Phase 2: Execution and feedback.} Based on the preprocessing phase outputs, \MainAgent enters an iterative refinement loop to progressively construct a functional Dockerfile capable of executing the target repository within the current hardware environment. 
This phase employs two complementary agents.
The \emph{WriteDocker Agent} generates or revises Dockerfile specifications based on the established plan and current execution state, while the \textit{RunDocker Agent} executes the generated Dockerfile within an isolated containerized environment and captures detailed execution logs and error diagnostics.

\vspace{.1cm}\noindent\textbf{Phase 3: Summarization and context management.} 
Following each execution cycle, the \emph{Summarize Agent} processes the execution feedback to synthesize the current Dockerfile completion status and formulate actionable modification recommendations. 
This agent maintains contextual continuity across iterations by organizing error information and corrective suggestions into structured feedback for subsequent refinement cycles. 
Additionally, it monitors success metric to determine loop termination conditions, ensuring efficient convergence to a functional environment configuration.

\vspace{.1cm}\noindent\textbf{Rule-based tool integration.} To optimize computational costs and enhance operational efficiency, \MainAgent strategically incorporates rule-based auxiliary tools alongside its intelligent agent framework. 
These deterministic components handle routine operational tasks that do not require sophisticated reasoning capabilities, thus reducing token consumption while maintaining system reliability.

The framework employs three primary rule-based utilities:
\begin{itemize}[leftmargin=*]
\item \textit{Workspace Initialization}, which systematically provisions the \MainAgent working directory structure and creates essential configuration files (e.g., \texttt{plan.txt}, \texttt{history.txt}, \texttt{status.txt}, \texttt{EnvAgent.dockerfile}, and hardware specification files) through file system operations;
\item \textit{Directory Scaffolding}, which establishes standardized folder hierarchies and dependency tracking mechanisms without requiring agent intervention; and
\item \textit{Workspace State Management}, which maintains execution context through automated log aggregation, iteration tracking, and status synchronization across the multi-agent workflow.
\end{itemize}

The goal of this hybrid architecture is to leverage rule-based automation for predictable tasks while reserving LLM capabilities for complex reasoning and decision-making processes, making the overall agentic workflow token-efficient and remain generalizable to diverse scenarios. 
Table~\ref{tab:EnvAgent_agents} presents a comprehensive overview of the EnvAgent framework's components and execution orders. 

\begin{table}[!t]
\centering
\setlength{\tabcolsep}{3pt}
\caption{\small Overview of \textbf{MainAgent}. 
    In Type column, \textbf{rule-based} means purely script-driven processing of the workspace, 
     \textbf{agent-based} means context-aware analysis and manipulation of the workspace by an agent.}
\label{tab:EnvAgent_agents}
\footnotesize

\resizebox{\columnwidth}{!}{%
\begin{tabular}{rlp{8cm}r}
\toprule
\textbf{Order} & \textbf{Component} & \textbf{Description} & \textbf{Type} \\
\midrule
\multicolumn{4}{c}{\textbf{Initialization Phase}} \\
\midrule
\circled{1} & Stats & Records session start time and system metrics for performance tracking & Rule \\
\circled{2} & Workspace Initial & Creates workspace structure with essential files (plan, status, etc.) & Rule \\
\midrule
\multicolumn{4}{c}{\textbf{Analysis \& Planning Phase}} \\
\midrule
\circled{3} & Scanning & Scans repository structure, builds file tree, and index environment files & Agent \\
\circled{4} & Hardware Checking & Detects hardware specifications using system introspection and LLM & Agent \\
\circled{5} & Planning & Synthesizes repository context to generate build strategy & Agent \\
\circled{6} & Hardware Adjustment & Refines the initial plan based on hardware limitations & Agent \\
\midrule
\multicolumn{4}{c}{\textbf{Iterative Execution Phase}} \\
\midrule
\circled{7a} & Docker Initial & Generates initial Dockerfile based on planning phase outputs & Agent \\
\circled{7b} & Docker Revision & Revises Dockerfile based on execution feedback and error analysis & Agent \\
\circled{8} & Run Dockerfile & Executes Dockerfile in containerized environment and captures logs & Rule \\
\circled{9} & Summarize & Analyzes execution results, identifies errors, and generates actionable recommendations for next iteration & Agent \\
\circled{10} & Logging & Aggregates current iteration status into disk and updates workspace & Rule \\
\circled{11} & Status Checking & Evaluates success criteria and end construction & Rule \\
\midrule
\multicolumn{4}{c}{\textbf{Finalization Phase}} \\
\midrule
\circled{12} & Stats & Records session end time and computes metrics & Rule \\
\circled{13} & Report & Displays formatted execution time and status & Rule \\
\bottomrule
\end{tabular}%
}
\end{table}

\subsection{\textsc{EnvBench} Curation}
To systematically evaluate automated environment configuration, we introduce
\textsc{EnvBench}, a benchmark spanning 54 carefully curated repositories from
two complementary sources: (1) reproducible research artifacts from the
Chameleon Trovi service~\cite{keahey2023three}, and (2) multi-lingual open-source software from Multi-SWE-bench, covering more than seven programming languages~\cite{zan2025multi}.
\MainAgent employs a three-dimensional framework tailored to environment construction assessment.

\begin{itemize}[leftmargin=*]
\item \emph{Basic Structure Verification} -- validates the presence of essential files, directories, and environment variables required for proper software operation;
\item \emph{Dependency Integrity Assessment} -- verifies not only dependency installation but also version compatibility and configuration correctness; and
\item \emph{Functional Validation} -- executes targeted commands to confirm that the environment supports the intended software functionality without requiring full application execution.
\end{itemize}

\begin{figure}[!t]
  \centering
  \includegraphics[width=\linewidth]{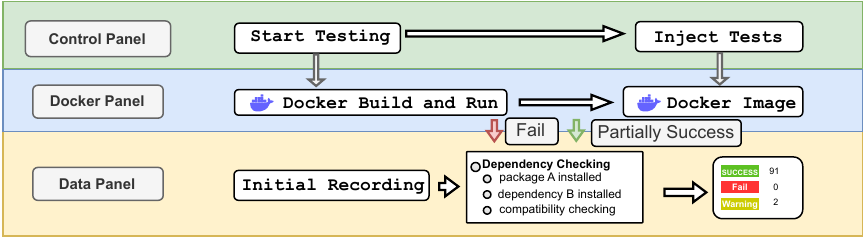}
\caption{\small Overview of \textsc{EnvBench}. Given a generated Dockerfile,
\textsc{EnvBench} builds and launches the corresponding container, injects
repository-specific declarative rubric checks, and executes them inside the
constructed environment. It records each check as pass, warning, or fail and
aggregates the outcomes into a report covering environment structure,
dependency integrity, and functional correctness.}
\end{figure}

Our evaluation design focuses on environment configuration rather than
comprehensive software testing. Because the objective is to verify that an
environment is correctly established, most checks do not require complete
application execution. Instead, \textsc{EnvBench} evaluates the Docker
containers produced from generated Dockerfiles by running environment-specific
validation commands and checking their outputs. Functionally equivalent
environments therefore receive consistent scores even when their
implementations differ.

\section{Experimental Setup}
\vspace{.1cm}\noindent\textbf{Datasets.}
Our benchmark comprises 54 heterogeneous repositories (Table~\ref{tab:dataset_overview}) covering a broad range of reproducibility requirements and application domains.
Among them, 21 come from Trovi hosted on Chameleon~\cite{trovi} with the label ``reproducible research''~\cite{keahey2020lessons}, so we can ensure the evaluation in the benchmark for environment setup is grounded. 
The remaining 33 artifacts come from Multi-SWE-bench~\cite{zan2025multi} covering popular real-world software beyond research prototypes, which further expand the benchmark diversity in terms of programming languages (e.g., Java, TypeScript, JavaScript, Go, Rust, C, and C++) and software functionalities. 
Such a diverse dataset can evaluate the agent across different programming paradigms, build systems, and deployment requirements, providing a comprehensive testbed for automated environment configuration capabilities.

\vspace{.1cm}\noindent\textbf{Baselines.}
We consider two widely used, representative agentic frameworks as our baselines: 
(1) \emph{Claude Code} based on Anthropic's reasoning-optimized models claude-opus-4 and claude-haiku-3.5, and
(2) \emph{Codex} based on OpenAI's code-specialized models gpt-4.1 and gpt-4.1-mini. 
We intentionally pick the cheap models to evaluate the delta brought by our agent harness, as powerful frontier models can often complement weak agentic architectures, making it unclear to evaluate the contribution of the agent framework.
Similar to EnvAgent, the baseline methods are provisioned with unrestricted access to the target repository and complete hardware inspection capabilities to ensure a fair comparison. 
All baselines and \MainAgent are initialized with an identical system prompt:
\begin{center}
\colorbox{lightgray}{
\begin{minipage}{\dimexpr\linewidth-2\fboxsep\relax}
\raggedright
\footnotesize
\textbf{System Prompt} \\
\texttt{Please take a careful look at the current hardware and repo, and create a new Dockerfile named \texttt{codex.dockerfile} (or \texttt{claude.dockerfile}) at the root of this repo that, when built, puts me in a \texttt{/bin/bash} CLI setting at the root of the repository, with the repository installed.}
\end{minipage}
}
\end{center}
This prompt instructs the model to inspect both the available hardware and the project repository, and to generate a new Dockerfile at the root of the repository. 
When built, the Dockerfile will include a working environment in which the repository is correctly installed and accessible from \texttt{/bin/bash} command-line interface at the repository root. 
This ensures that each system begins with the same context and a consistent environment setup procedure critical for reproducibility and fair evaluation.
\subsection{Evaluation Rubrics}
To systematically evaluate \MainAgent's environment construction capabilities, we develop \EnvBench --- a benchmark and, more importantly, a \emph{general, rubric-driven evaluation framework} that targets the often-overlooked problem of verifying whether an execution environment has been correctly \emph{constructed}, rather than whether software features are correct.
Unlike ordinary unit tests, which presuppose a working environment, \EnvBench checks the environment itself; and unlike prior efforts that rely on bespoke, per-repository validation scripts, it lets users specify only a declarative \emph{rubric} while a shared engine performs the actual checking.
The rubric defines a set of fine-grained requirements that a correctly constructed environment should satisfy, while the shared evaluation engine executes these checks and aggregates their outcomes into a quantitative environment-quality score.
We detail the rubric design, its taxonomy of checks, and the evaluation procedure below.
\begin{table}[!t]
\centering
\setlength{\tabcolsep}{3pt}
\caption{\small EnvAgent Dataset: Repository Overview}
\label{tab:dataset_overview}
\footnotesize
\resizebox{\columnwidth}{!}{%
\begin{tabular}{lll|lll|lll}
\toprule
\textbf{Repo} & \textbf{Lang} & \textbf{Tests} &
\textbf{Repo} & \textbf{Lang} & \textbf{Tests} &
\textbf{Repo} & \textbf{Lang} & \textbf{Tests} \\
\midrule
\multicolumn{9}{c}{\textbf{Chameleon Trovi (21 repos)}} \\
\midrule
Reltr & Python & 71 & Lottory & Python & 97 & Seed-gnn & Python & 73 \\
Tabpfn & Python & 69 & Rsnn & Python & 84 & P4ctl & Python & 110 \\
Crossprefetch & C/C++ & 67 & Symmc & Java & 40 & Fairify & Java & 72 \\
Exli & Java & 49 & Sixthsense & Python & 64 & Probfuzz & Python & 104 \\
Gluetest & Python & 82 & Flex & Python & 73 & Acto & Python & 63 \\
Baleen & C/C++ & 63 & Silhouette & C/C++ & 74 & Anvil & Rust & 70 \\
Elect & C/C++ & 45 & Rfuse & C/C++ & 48 & Metis & C/C++ & 104 \\
\midrule
\multicolumn{9}{c}{\textbf{Multi-SWE-bench (33 repos)}} \\
\midrule
Zstd & C/C++ & 71 & Jq & C/C++ & 95 & Catch2 & C/C++ & 39 \\
Fmt & C/C++ & 105 & Json & C/C++ & 96 & Simdjson & C/C++ & 59 \\
Cpp-httplib & C/C++ & 41 & Cli & Go & 72 & Grpc-go & Go & 104 \\
Go-zero & Go & 34 & Fastjson2 & Java & 52 & Logstash & Java & 42 \\
Mockito & Java & 101 & Github-readme-stats & JS & 59 & Axios & JS & 76 \\
Express & JS & 50 & Dayjs & JS & 101 & Insomnia & JS & 104 \\
Svelte & JS & 74 & Ripgrep & Rust & 82 & Clap & Rust & 59 \\
Nushell & Rust & 109 & Serde & Rust & 65 & Bat & Rust & 72 \\
Fd & Rust & 58 & Rayon & Rust & 81 & Bytes & Rust & 60 \\
Tokio & Rust & 60 & Tracing & Rust & 66 & Darkreader & TS & 26 \\
Material-ui & TS & 32 & Core & TS & 42 & Ponyc & C/C++ & 74 \\
\bottomrule
\end{tabular}%
}
\end{table}

\vspace{.1cm}\noindent\textbf{From bespoke scripts to declarative rubrics.}
A key obstacle to environment-focused evaluation is that, unlike ordinary unit tests which presuppose a working environment, validating environment \emph{construction} has traditionally required hand-written, repository-specific shell scripts---tedious to author and hard to compare across projects.
To remove this burden, we design \EnvBench as a \emph{general, declarative evaluation framework}: the user specifies only a \emph{rubric}---a machine-readable description of what a correctly configured environment must satisfy---while a shared execution engine performs the checking.
This cleanly separates \emph{what} to verify (the rubric, the only artifact the user authors) from \emph{how} to verify it (reusable checker primitives supplied by the framework), so extending the benchmark to a new repository reduces to writing one rubric rather than a validation program.

\vspace{.1cm}\noindent\textbf{Rubric specification.}
A rubric is a list of fine-grained checks, where each check declares a reusable checker primitive (its \texttt{type}), the primitive's parameters, a score weight, a semantic \texttt{category}, and optional prerequisite checks (\texttt{requires}) that impose execution ordering.
Each check evaluates one concrete property of the constructed environment, such as whether a required file or directory exists, whether a required executable is available, whether an environment variable is correctly set, or whether a setup, build, or test command succeeds.
Seven primitives suffice to express every check in our suite: \texttt{run\_command} (a shell command must exit successfully), \texttt{commands\_exist} (executables are on \texttt{PATH}), \texttt{files\_exist}/\texttt{dirs\_exist} (paths are present), \texttt{file\_contains}/\texttt{output\_contains} (a file, or a command's output, contains required content), and \texttt{envvar\_set} (an environment variable is set).
Because these primitives are declarative and composable, authoring a rubric requires no scripting.

\vspace{.1cm}\noindent\textbf{Two-level taxonomy.}
Each check is organized along a two-level taxonomy (Table~\ref{tab:envbench_taxonomy}).
The first level is a \emph{validation category} capturing which aspect of environment readiness is probed: \textit{Structure} (the repository layout and declared dependencies are in place), \textit{Configuration} (tools, interpreters, versions, and environment variables are correctly installed), and \textit{Functionality} (build, test, and run commands actually execute successfully), together with a minimal \textit{Integration} probe for end-to-end container startup.
The second level is the checker primitive that realizes the check.
Table~\ref{tab:envbench_taxonomy} reports the full cross-classification with per-cell counts over the rubric suite.
\begin{table}[!t]
\centering
\caption{\small \EnvBench rubric taxonomy. Each check is classified by a first-level validation \emph{category} (which aspect of environment readiness it probes) and realized by a second-level reusable \emph{checker primitive} (\texttt{type}). Counts are aggregated over the rubric suite (2{,}108 checks).}
\label{tab:envbench_taxonomy}
\footnotesize
\setlength{\tabcolsep}{3pt}
\renewcommand{\arraystretch}{1.2}
\resizebox{\columnwidth}{!}{%
\begin{tabular}{lllr}
\toprule
\textbf{Category (L1)} & \textbf{Checker primitive (L2)} & \textbf{Each check asserts} & \textbf{\#} \\
\midrule
\multirow{4}{*}{\textbf{Structure} (579)}
 & \texttt{files\_exist}     & required files are present               & 227 \\
 & \texttt{dirs\_exist}      & required directories are present         & 145 \\
 & \texttt{file\_contains}   & a file pins required content             & 123 \\
 & \texttt{run\_command}     & structural probe command succeeds        & 84 \\
\midrule
\multirow{5}{*}{\textbf{Configuration} (863)}
 & \texttt{run\_command}     & install/setup command succeeds           & 440 \\
 & \texttt{commands\_exist}  & required tools are on \texttt{PATH}      & 227 \\
 & \texttt{output\_contains} & command output matches (e.g., version)   & 118 \\
 & \texttt{envvar\_set}      & environment variable is set              & 62 \\
 & \textit{other}            & file/dir presence checks                 & 16 \\
\midrule
\multirow{3}{*}{\textbf{Functionality} (665)}
 & \texttt{run\_command}     & build/test/run command exits 0           & 630 \\
 & \texttt{output\_contains} & program output matches expected          & 17 \\
 & \textit{other}            & artifact presence checks                 & 18 \\
\midrule
\textbf{Integration} (1) & \texttt{run\_command} & end-to-end minimal container run & 1 \\
\midrule
\textbf{Total} & \textbf{7 primitives} & & \textbf{2{,}108} \\
\bottomrule
\end{tabular}%
}
\end{table}

\vspace{.1cm}\noindent\textbf{Execution and scoring.}
Evaluation proceeds in three automated stages.
First, the rubric is injected into the \MainAgent workspace.
Second, the generated Dockerfile is built into a container.
Third, the evaluation engine executes every rubric check inside the resulting container while honoring \texttt{requires} dependencies.
Each check produces one of three outcomes: \textsc{pass}, \textsc{fail}, or \textsc{warning}.
These outcomes are converted into numerical scores and aggregated according to the check weights to obtain a repository-level rubric score.
Specifically, we assign a normalized outcome score of $1$ to \textsc{pass}, $0$ to \textsc{fail}, and $\alpha$ to \textsc{warning}, where $\alpha\in[0,1]$ is a fixed partial-credit value.
For repository $r_i$ with $M_i$ rubric checks, check $k$ has weight $w_{ik}$ and outcome score $s_{ik}$.
The resulting rubric-based environment quality score is therefore ${S_{TSSR}}_i = \frac{\sum_{k=1}^{M_i} w_{ik}s_{ik}}{\sum_{k=1}^{M_i} w_{ik}}$.
Thus, the rubric defines \emph{what} is evaluated, while $S_{TSSR}$ quantitatively summarizes \emph{how well} the constructed environment satisfies those requirements.
In parallel, successful Docker build and runtime execution determines whether the environment is executable, which is used to compute $S_{ECSR}$.
The two metrics capture complementary aspects of environment construction and are combined into the final composite score described in Section~\ref{sec:EV}.
This framework turns benchmark curation into rubric authoring: across the \EnvBench suite, the rubrics encode 2{,}108 weighted checks spanning three primary validation categories, one integration probe, and seven checker primitives (Table~\ref{tab:envbench_taxonomy}).
Together, these checks provide a consistent and reusable evaluation interface for assessing \MainAgent's ability to construct environments across diverse software ecosystems and build complexities.
Table~\ref{tab:dataset_overview} summarizes the repositories covered by \EnvBench.
\subsection{Evaluation Metrics}
\label{sec:EV}
To comprehensively assess \MainAgent's performance in automated environment configuration, we employ complementary metrics that capture two distinct aspects of environment construction: (1) whether the agent can successfully construct an executable environment, and (2) how well the resulting environment satisfies the rubric-defined requirements.
We further combine these two aspects into a single overall score.

\vspace{.1cm}\noindent\textbf{Executable count.}
This metric measures how many functional Docker images \MainAgent successfully generates that build and run without critical errors on the target hardware.
For a given repository $r_i$ and hardware configuration $h_j$, we define $\text{executable}_{i,j}$ as a binary indicator that equals 1 if the generated Dockerfile builds and runs successfully, and 0 otherwise.
The total number of executable environments across the dataset is $\text{executable} = \sum_{i=1}^{N} \text{executable}_{i,j}$, where $N$ represents the total number of repositories in the evaluation dataset.

\vspace{.1cm}\noindent\textbf{Environment Construction Success Rate ($S_{ECSR}$).}
Besides the total count, we report the proportion of repositories for which \MainAgent successfully constructs an executable environment: $S_{ECSR} = \frac{1}{N}\sum_{i=1}^{N}\text{executable}_{i,j}$.
$S_{ECSR}$ measures the agent's fundamental ability to produce a viable environment, independently of the fine-grained rubric score.

\vspace{.1cm}\noindent\textbf{Test Suite Success Rate ($S_{TSSR}$).}
For each executable repository, $S_{TSSR}$ measures the quality of the constructed environment based on the weighted outcomes of its \EnvBench rubric checks.
For repository $r_i$, let $M_i$ denote the number of rubric checks, $w_{ik}$ the weight assigned to check $k$, and $s_{ik}$ its normalized outcome score.
We define the outcome score as $s_{ik}=1$ for \textsc{pass}, $s_{ik}=\alpha$ for \textsc{warning}, and $s_{ik}=0$ for \textsc{fail}, where $\alpha\in[0,1]$ is the fixed partial-credit value assigned to warnings.
The repository-level $S_{TSSR}$ is then ${S_{TSSR}}_i = \frac{\sum_{k=1}^{M_i}w_{ik}s_{ik}}{\sum_{k=1}^{M_i}w_{ik}}$.
This formulation ensures that more important environment requirements contribute proportionally more to the final score, while warnings receive partial credit rather than being treated as either complete success or complete failure.
Let $\mathcal{R}=\{i\mid\text{executable}_{i,j}=1\}$ denote the subset of repositories with executable environments.
We aggregate $S_{TSSR}$ over this executable subset as the weighted mean of repository-level rubric scores: ${S_{TSSR}}_{\mathcal{R}}=\frac{1}{|\mathcal{R}|}\sum_{i\in\mathcal{R}}{S_{TSSR}}_i$.
Thus, $S_{TSSR}$ evaluates environment quality conditional on successful execution, while $S_{ECSR}$ accounts for repositories for which an executable environment could not be constructed.

\vspace{.1cm}\noindent\textbf{Final Composite Score (FCS).}
To summarize both environment construction capability and environment quality in a single scalar, we compute $\text{FCS}=\left(\frac{S_{ECSR}}{100}\right)\times{S_{TSSR}}_{\mathcal{R}}$.
Here, $S_{ECSR}$ is expressed as a percentage, whereas ${S_{TSSR}}_{\mathcal{R}}$ is normalized to $[0,1]$.
FCS therefore captures both the likelihood that a repository has an executable environment and the quality of that environment conditional on executability.
If no repository is executable (i.e., $|\mathcal{R}|=0$), we define $\text{FCS}=0$.
Together, these metrics provide complementary views of \MainAgent's effectiveness.
Executable count reports the absolute number of viable environments, $S_{ECSR}$ normalizes construction success across datasets, ${S_{TSSR}}_{\mathcal{R}}$ measures the quality of executable environments using the weighted rubric checks, and FCS integrates environment construction success and environment quality into a single overall measure.
In this formulation, the rubric is the fine-grained evaluation specification, the individual check outcomes provide the underlying evaluation signals, and the metrics aggregate these signals into standardized measures for quantitative comparison across repositories and environments.
\subsection{Implementation}
We consider a diverse set of LLMs to ensure a comprehensive evaluation
(Table~\ref{tab:envagent_results_all}).
To enable systematic comparison and streamline the model management, we develop a unified API management framework using Forge~\cite{tensorblock_forge}, which provides centralized model orchestration and usage analytics.

\vspace{.1cm}\noindent\textbf{Unified interface and configuration management.} 
Our framework abstracts model-specific implementations through a standardized interface layer, enabling seamless model switching across all agent components without code modification. 
Model selection and parameters are managed through a declarative configuration system that supports both global and component-specific model assignments. 
Researchers can easily configure experiments to compare model performance across the entire \MainAgent pipeline or evaluate specific agents with different models while maintaining consistency in other components. 
This design ensures that complex multi-agent workflows can be executed with consistent parameters while maintaining model-agnostic operation logic, and provides the flexibility needed for systematic ablation studies and model comparison experiments with minimal configuration overhead.

\vspace{.1cm}\noindent\textbf{Centralized usage tracking.} The Forge-based management system provides comprehensive usage analytics, including token consumption, API call frequencies, and cost tracking across all models and agent components.
This centralized monitoring enables precise resource accounting and facilitates fair comparison between different model configurations. The system automatically logs detailed metrics for each agent invocation, supporting both per-iteration and aggregate analysis of model utilization patterns.
This unified model management approach ensures experimental reproducibility while providing the flexibility needed for comprehensive model evaluation across the complex multi-agent \MainAgent workflow.

\vspace{.1cm}\noindent\textbf{Infrastructure and implementation architecture.} 
We conduct all experiments on the Chameleon Cloud testbed using single-node servers configured with x86\_64 architecture running Ubuntu 20.04 LTS. 
For repositories requiring GPU acceleration, we provision NVIDIA A100 GPUs to ensure compatibility with compute-intensive workloads. 
We implement \MainAgent from the ground up and evaluate every method on
the same pre-provisioned hardware for a controlled comparison. A separate
preliminary study of settings without pre-provisioned hardware is reported in
Section~\ref{subsec:moving-to-cloud}.
\section{Evaluation}
\label{sec:eval}

We aim to answer the following research questions:

\begin{itemize}[leftmargin=*]
    \item \textbf{RQ1:} How effective is \MainAgent at generating executable Docker environments compared to baseline approaches?
    \item \textbf{RQ2:} What is the overall quality and functional correctness of environments generated by \MainAgent as measured by comprehensive benchmarks?
\end{itemize}

\subsection{RQ1: Environment Generation Effectiveness}
\label{subsec:rq1}

As shown in Table~\ref{tab:envagent_results_all}, \MainAgent produces more executable environments than both baselines across matched model families. 
Its executable rates range from 66.7\% to 90.7\%, while the baselines achieved 37.0\% to 51.9\%, respectively.
Note that executability is a prerequisite for this task, so if a Docker image cannot be built and run, downstream dependency and functional checks cannot be applied, and the agent receives a score of zero.

For the resource efficiency, e.g., tokens, time, dollar cost, etc., in Table~\ref{tab:envagent_results_all},
\textit{All} reports the mean over all 54 repository runs for each agent--model configuration, including unsuccessful attempts, whereas \textit{Success} reports the conditional mean over only the runs that produced an executable environment. 
Comparing these columns helps distinguish the overall evaluation cost from the resources consumed by successful constructions. 
In particular, failed runs may terminate early and lower the \textit{All} average, while successful runs typically include the full build, execution, and validation workflow.
The higher executable counts come with increased token consumption.

\begin{table}[!t]
\centering
\small
\setlength{\tabcolsep}{3pt}
\renewcommand{\arraystretch}{1.25}

\caption{\textsc{EnvAgent} results across 54 repositories per agent--model
configuration. For tokens, time, and cost, \textit{All} is the mean over all
runs, whereas \textit{Success} is the mean over runs that produced an
executable environment. Input/output tokens are measured in thousands, cost is
in dollars, and FCS (Final Composite Score) is in \%.}

\resizebox{\columnwidth}{!}{%
\begin{tabular}{cc|cc|cc|cc|cc|cc|c}
\toprule
\multirow{2}{*}{\textbf{Agent}} &
\multirow{2}{*}{\textbf{Model}} &
\multicolumn{2}{c|}{\textbf{Executable}} &
\multicolumn{2}{c|}{\textbf{Input Tokens}} &
\multicolumn{2}{c|}{\textbf{Output Tokens}} &
\multicolumn{2}{c|}{\textbf{Time}} &
\multicolumn{2}{c|}{\textbf{Cost}} &
\multirow{2}{*}{\textbf{FCS}} \\
& &
\textbf{Count} & \textbf{Rate (\%)} &
\textbf{All} & \textbf{Success} &
\textbf{All} & \textbf{Success} &
\textbf{All} & \textbf{Success} &
\textbf{All} & \textbf{Success} &
\\
\midrule

\multirow{2}{*}{codex}
& gpt-4.1
& 21 & 38.9
& 376.6 & 418.1
& 0.3 & 0.2
& 0:00:46 & 0:00:46
& 0.75 & 0.83
& 26.15 \\

& gpt-4.1-mini
& 28 & 51.9
& 331.2 & 282.6
& 0.3 & 0.3
& 0:00:39 & 0:00:33
& 0.13 & 0.11
& 22.31 \\

\midrule

\multirow{2}{*}{claude-code}
& claude-opus-4
& 25 & 46.3
& 26.6 & 24.7
& 0.3 & 0.2
& 0:01:24 & 0:01:15
& 0.58 & 0.58
& 30.10 \\

& claude-haiku-3.5
& 20 & 37.0
& 18.6 & 17.9
& 0.2 & 0.2
& 0:00:23 & 0:00:23
& 0.02 & 0.02
& 18.21 \\

\midrule

\multirow{7}{*}{EnvAgent}
& gpt-4.1
& 48 & 88.9
& 465.2 & 375.9
& 105.5 & 100.2
& 0:32:30 & 0:27:28
& 2.66 & 2.33
& 70.07 \\

& gpt-4.1-mini
& 36 & 66.7
& 368.4 & 253.7
& 90.2 & 95.6
& 0:31:36 & 0:28:27
& 0.58 & 0.51
& 53.41 \\

& claude-opus-4
& 49 & 90.7
& 322.8 & 294.1
& 75.5 & 79.9
& 0:42:55 & 0:45:01
& 10.5 & 10.1
& 74.01 \\

& claude-haiku-3.5
& 41 & 75.9
& 272.3 & 184.3
& 15.2 & 14.7
& 0:11:39 & 0:10:43
& 0.27 & 0.21
& 50.59 \\

& deepseek-r1
& 30 & 55.6
& 351.9 & 248.6
& 104.7 & 76.6
& 1:42:10 & 0:57:30
& 0.40 & 0.29
& 42.25 \\

& deepseek-v3
& 39 & 72.2
& 319.1 & 278.4
& 42.2 & 44.4
& 0:27:21 & 0:27:33
& 0.13 & 0.13
& 49.54 \\

& gemini-2.5-pro
& 35 & 64.8
& 556.2 & 612.1
& 198.9 & 239.6
& 0:50:39 & 0:54:07
& 2.68 & 3.36
& 66.81 \\

\bottomrule
\end{tabular}%
}
\label{tab:envagent_results_all}
\end{table}

\MainAgent allocates additional reasoning budget to its
plan--execute--feedback--revise loop (Section~\ref{sec:EA}), including multi-round error localization and Dockerfile repair. 
Baselines often terminate earlier after a failed attempt and therefore consume fewer tokens, but also construct fewer executable environments. 
Across model families, stronger reasoning models (e.g., GPT-4.1 and Claude Opus series) achieve more executable environments, whereas smaller models offer lower cost and latency.
These observations motivate a possible curriculum policy that first uses a lower-cost model and escalates only after an unsuccessful attempt.

Longer runs are often dominated by compilation and image-building costs rather than agent inference alone. 
The standardized evaluation harness records the same inputs and measurements for every method. Build caching and artifact reuse could thus reduce both latency and cost in future implementations.

As shown in Table~\ref{tab:envagent_results_all}, \MainAgent produces
36--49 executable environments across the four matched model settings, while
the baselines produce 20--28. Averaged over these settings, the corresponding
counts are 43.5 and 23.5, respectively.

\subsection{RQ2: Environment Quality and Functional Correctness}
\label{subsec:rq2}

We report \textsc{EnvBench} composite scores (defined in
Section~\ref{sec:EV}) across agents and models. As shown in
Table~\ref{tab:envagent_results_all}, \MainAgent achieves an average FCS
of 62.02 across the four matched model settings, compared with 24.19 for the
baselines---a margin of 37.83 percentage points. The improvement is consistent
within each matched model family: FCS increases from 26.15 to 70.07 with
GPT-4.1, from 22.31 to 53.41 with GPT-4.1-mini, from 30.10 to 74.01 with
Claude Opus 4, and from 18.21 to 50.59 with Claude Haiku 3.5. Thus, the gains
are not attributable to a single model provider or model scale.

The strongest \MainAgent configuration reaches an FCS of 74.01, while
even its weakest configuration among all seven evaluated models reaches 42.25,
exceeding the strongest baseline at 30.10. This distinction is important
because FCS jointly captures whether an environment is executable and how well
an executable environment satisfies the weighted \textsc{EnvBench} rubrics.
Together with the executable-count results in RQ1, these scores show that the
gains persist under an end-to-end metric that evaluates rubric satisfaction
rather than stopping at the initial build stage.

\subsection{Moving to Cloud Environment}
\label{subsec:moving-to-cloud}

In this paper, the main evaluation assumes that a machine with fixed hardware is already available, so the effort is focused on constructing the environment on the given machine. 
As a preliminary effort, we also briefly explore the setting when an experiment must run on Cloud environment where the required hardware has not yet been provisioned~\cite{keahey_2025_15306610,keahey2023three}. 
In this study, we deliberately obtained 20 additional repositories from Trovi with heterogeneous hardware requirements and is separate from the main 54-repository \textsc{EnvBench} evaluation.
The goal of this study is to understand whether we can easily extend our 

\begin{figure}[!t]
\centering
\includegraphics[width=\linewidth]{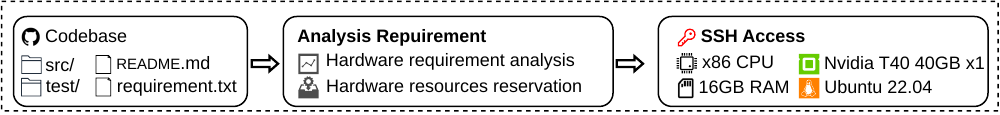}
\caption{\small Preliminary repository-based hardware provisioning for cloud experiments. 
The information in the repository is used to infer hardware requirements to provision, which are then matched against the provisioned cloud resources before SSH access is returned.}
\label{fig:cloud_provisioning}
\end{figure}

As Figure~\ref{fig:cloud_provisioning} shows, an agent infers CPU, memory, storage, operating-system, and accelerator requirements from repository evidence, matches them to available Chameleon resources, and returns SSH access after reservation~\cite{keahey2020lessons,mei2024aios}. Representative cases are reported in Table~\ref{tab:hw_provisioning}. 
As this study is preliminary, we do not treat the workflow as a another system or include it in the main \MainAgent--\textsc{EnvBench} comparison.
\section{Related Work}

\vspace{.1cm}\noindent\textbf{Agentic coding benchmarks.}
Recent benchmarks evaluate code agents on software engineering tasks including
automated bug repair~\cite{liu2024marscode,chen2023teaching,jimenez2023swe,
kang2025explainable,zhou2025large}, code-efficiency
optimization~\cite{lin2025eco,chen2021evaluating,
shetty2025gsochallengingsoftwareoptimization}, code migration and
transformation~\cite{cheng2025codemenv,pomian2024assist}, and penetration
testing~\cite{gioacchini2024autopenbench,david2025multi,jing2024secbench,
lee2025sec,zhang2024cybench,guo2025repoaudit,peng2025cweval}.
Building these benchmarks often requires manually constructing executable
Docker environments~\cite{swe-smith-manual}, which must be maintained as
dependencies evolve, limiting scalability and potentially excluding repositories
with complex or changing build requirements~\cite{rahman2024impact}.
\textsc{EnvAgent} addresses this infrastructure bottleneck by automating
repository-level environment construction.

\vspace{.1cm}\noindent\textbf{Repo-level execution environment construction.}
The reproducibility crisis in computational research has motivated efforts to automate
environment setup and result reproduction~\cite{gundersen2018state}.
Container-based solutions such as Docker improve portability, but creating
reliable Dockerfiles still requires substantial effort from domain
experts~\cite{cito2017empirical}. Existing LLM-based approaches provide partial
support for dependency installation and configuration, but do not systematically
combine iterative construction with rubric-driven environment
evaluation~\cite{hou2024large}.
Research reproduction benchmarks, such as PaperBench~\cite{starace2025paperbench},
focus primarily on reproducing computational experiments rather than systematic
environment configuration assessment.
Traditional reproducibility tools, such as ReproZip~\cite{chirigati2016reprozip}
and CDE~\cite{guo2018cde}, can capture computational requirements but require
manual intervention for complex dependency chains~\cite{chirigati2016reprozip,guo2018cde}.

\vspace{.1cm}\noindent\textbf{Code agents.}
LLM-based software engineering agents have rapidly evolved to automate a broad
range of development tasks~\cite{liu2024marscode,kumar2025sharp,
chen2025unveiling,tufano2024autodev,wang2024openhands,hong2024metagpt,
qian2023chatdev,grishina2025fully,liang2024large,li2025editlord,
chen2024advancing}.
Notably, Claude Code~\cite{claude-code} and Codex~\cite{codex} provide agentic coding capabilities that run directly in terminals, supporting end-to-end workflows including feature implementation from natural language descriptions, debugging through error analysis, and codebase navigation with global project awareness.

However, most code agents rely on an already executable environment for
test-driven development and execution-based evaluation~\cite{yang2024swe,
shetty2025gsochallengingsoftwareoptimization}. This assumption restricts their
applicability to repositories whose environments can be prepared
reliably~\cite{rahman2024impact}. \textsc{EnvAgent} is complementary: it
constructs the execution environment needed to extend agent training and
evaluation to a broader set of repositories.

\section{Threats to Validity}

\vspace{.1cm}\noindent\textbf{Limited evaluated models and baselines.} Our
evaluation covers seven model configurations, including GPT-4.1, Claude Opus 4,
DeepSeek-R1, and Gemini-2.5-Pro
(Table~\ref{tab:envagent_results_all}). These models may not represent the
performance of future model families. Our baseline comparison is also limited
to two general-purpose code agents, Codex and Claude Code, because we are not
aware of a directly comparable agent specialized in end-to-end environment
construction. Consequently, the measured improvements should not be
generalized to every model or agent architecture.

\vspace{.1cm}\noindent\textbf{Increased resource consumption.}
\textsc{EnvAgent}'s plan--execute--feedback--revise loop consumes substantially
more tokens and wall-clock time than the baselines in several configurations.
This cost may limit practical adoption even when the resulting environment is
more reliable. Build caching, artifact reuse, model cascading, and more
efficient error localization could reduce this overhead.

\vspace{.1cm}\noindent\textbf{Dockerfile-centric environment definition.} Our
framework defines environment construction success based on whether a single-container
Dockerfile satisfies the declarative rubric checks in \textsc{EnvBench}. This
definition may not fully capture real-world systems that require multi-node
deployments, distributed architectures, external services, or orchestration
platforms beyond a single container. While this choice is consistent with common
practices in SWE-related benchmarks and the widespread use of Dockerfiles for
environment specification, our preliminary cloud study only explores hardware
provisioning and does not address these multi-container and orchestration
requirements.

\section{Conclusion}

Execution environment construction remains a bottleneck in scaling the
training and evaluation of code agents across real-world repositories.
We present \textsc{EnvAgent}, a multi-agent framework that analyzes a
repository, plans its dependencies, and iteratively constructs and repairs a
Dockerfile using execution feedback. We also present \textsc{EnvBench}, a
rubric-driven benchmark covering 54 open-source repositories and reproducible
research artifacts. \textsc{EnvAgent} constructs an average of 43.5 executable
environments across four matched model settings, compared with 23.5 for Codex
and Claude Code, and achieves an average FCS of 62.02 versus 24.19. We
additionally report a preliminary study of repository-driven hardware
provisioning when a cloud machine is not yet available, separate from the core
\textsc{EnvAgent}--\textsc{EnvBench} contribution. We will release
\textsc{EnvAgent}, \textsc{EnvBench}, and the evaluation artifacts to support
future research on reproducible code-agent infrastructure.

\section*{Acknowledgment}
We thank the anonymous reviewers for their constructive comments and feedback, which helped us improve this paper. We also thank Maharani Irawan (University of Chicago) for early discussions on the definitions and boundaries of agents and tools. We are grateful to Professors Yiling Lou and Lingming Zhang (University of Illinois Urbana-Champaign) and Professor Claire Le Goues (Carnegie Mellon University) for valuable discussions on the design principles and perspectives of agents and software engineering. This work was supported by the National Science Foundation (Award No.~2431425), the Chameleon testbed~\cite{keahey2020lessons}, the TensorBlock Researcher Access Program, and the OpenAI Researcher Access Program. Generative models were used only to assist with language editing of this manuscript and the development of implementation code. They were not used to develop the research idea, formulate the problem, design the experiments, analyze the results, or make other substantive research decisions. All AI-assisted text and code were reviewed by the authors.

\bibliographystyle{IEEEtranN}

\bibliography{refs/main}
\newpage
\appendices

\section{Predicted vs.\ Provisioned Hardware Resources}
\label{app:hardware}

Table~\ref{tab:hw_provisioning} reports, for a representative subset of repositories, the hardware requirements specified by repository analysis together with the corresponding requirements inferred by the baseline agents (Codex and Claude) and the resources ultimately reserved by EnvAgent+ on bare-metal (TACC) and KVM platforms. These per-repository results complement the aggregate provisioning results discussed in Section~III-C.

\begin{table}[!h]
\centering
\caption{Specified and provisioned hardware resources by repository and resource category.}
\label{tab:hw_provisioning}
\resizebox{\columnwidth}{!}{%
\begin{tabular}{llccccc}
\toprule
Repository & Category & Specified & Codex & Claude & Bare metal (TACC) & KVM actual \\
\midrule
\multirow{5}{*}{\texttt{tao\_tutorials}} & CPU & 8 & 4 & 8 & 24 & 8 \\
 & RAM (GB) & 16 & 16 & 32 & 512 & 16 \\
 & Disk (GB) & 100 & 50 & 100 & 6400 & 160 \\
 & GPU & Yes & Yes & Yes & 4 x NVIDIA MI100 & not provisioned \\
 & GPU memory (GB) & 16 & 8 & 16 & 4 & not provisioned \\
\midrule
\multirow{5}{*}{\texttt{Gym}} & CPU & unknown & 4 & 16 & 48 & 4 \\
 & RAM (GB) & 8 & 16 & 64 & 384 & 8 \\
 & Disk (GB) & 5 & 20 & 100 & 6400 & 80 \\
 & GPU & No & Yes & Yes & N/A & N/A \\
 & GPU memory (GB) & unknown & 8 & 40 & N/A & N/A \\
\midrule
\multirow{5}{*}{\texttt{ai-avatar}} & CPU & unknown & 4 & 4 & 48 & 2 \\
 & RAM (GB) & unknown & 8 & 8 & 384 & 4 \\
 & Disk (GB) & unknown & 10 & 20 & 6400 & 40 \\
 & GPU & No & Yes & Yes & N/A & N/A \\
 & GPU memory (GB) & unknown & 4 & 8 & N/A & N/A \\
\midrule
\multirow{5}{*}{\texttt{Kilosort4}} & CPU & 4 & 4 & 8 & 16 & 4 \\
 & RAM (GB) & 8 & 16 & 32 & 192 & 8 \\
 & Disk (GB) & unknown & 10 & 50 & 4000 & 80 \\
 & GPU & Yes & Yes & Yes & 2 x NVIDIA K80 & not provisioned \\
 & GPU memory (GB) & 8 & 8 & 8 & 2 & not provisioned \\
\midrule
\multirow{5}{*}{\texttt{CLM-GS}} & CPU & 8 & 4 & 16 & 20 & 2 \\
 & RAM (GB) & 128 & 16 & 64 & 256 & 4 \\
 & Disk (GB) & 500 & 50 & 100 & 4000 & 40 \\
 & GPU & Yes & Yes & Yes & 2 x NVIDIA M40 & not provisioned \\
 & GPU memory (GB) & 24 & 8 & 24 & 2 & not provisioned \\
\midrule
\multirow{5}{*}{\texttt{CellViT-plus-plus}} & CPU & 16 & 4 & 16 & 20 & 16 \\
 & RAM (GB) & 32 & 16 & 64 & 256 & 32 \\
 & Disk (GB) & 30 & 20 & 100 & 4000 & 320 \\
 & GPU & Yes & Yes & Yes & 2 x NVIDIA M40 & not provisioned \\
 & GPU memory (GB) & 24 & 8 & 24 & 2 & not provisioned \\
\midrule
\multirow{5}{*}{\texttt{CellViT-Inference}} & CPU & 16 & 4 & 8 & 28 & 16 \\
 & RAM (GB) & 32 & 8 & 32 & 256 & 32 \\
 & Disk (GB) & 30 & 10 & 50 & 4000 & 320 \\
 & GPU & Yes & Yes & Yes & 2 x NVIDIA P100 & not provisioned \\
 & GPU memory (GB) & 24 & 8 & 16 & 2 & not provisioned \\
\midrule
\multirow{5}{*}{\texttt{plonky2-gpu}} & CPU & 8 & 4 & 16 & 24 & 8 \\
 & RAM (GB) & 16 & 8 & 64 & 512 & 16 \\
 & Disk (GB) & unknown & 10 & 50 & 6400 & 160 \\
 & GPU & Yes & Yes & Yes & 4 x NVIDIA MI100 & not provisioned \\
 & GPU memory (GB) & 12 & 4 & 16 & 4 & not provisioned \\
\midrule
\multirow{5}{*}{\texttt{svrtk-docker-gpu}} & CPU & 6 & 4 & 8 & 48 & 8 \\
 & RAM (GB) & 16 & 8 & 32 & 384 & 16 \\
 & Disk (GB) & 30 & 20 & 50 & 6400 & 160 \\
 & GPU & Yes & Yes & Yes & N/A & not provisioned \\
 & GPU memory (GB) & 12 & 4 & 8 & N/A & not provisioned \\
\midrule
\multirow{5}{*}{\texttt{node0}} & CPU & unknown & 4 & 4 & 48 & 16 \\
 & RAM (GB) & 32 & 8 & 8 & 384 & 32 \\
 & Disk (GB) & 80 & 20 & 50 & 6400 & 320 \\
 & GPU & Yes & No & No & N/A & not provisioned \\
 & GPU memory (GB) & 16 & 0 & 0 & N/A & not provisioned \\
\midrule
\multirow{5}{*}{\texttt{SciDOCX}} & CPU & 8 & 2 & 4 & 48 & 8 \\
 & RAM (GB) & 16 & 4 & 16 & 384 & 16 \\
 & Disk (GB) & 10 & 10 & 20 & 6400 & 160 \\
 & GPU & Yes & No & No & N/A & not provisioned \\
 & GPU memory (GB) & 6 & 0 & 0 & N/A & not provisioned \\
\midrule
\multirow{5}{*}{\texttt{U-Time}} & CPU & 4 & 4 & 8 & 48 & 4 \\
 & RAM (GB) & 8 & 8 & 32 & 384 & 8 \\
 & Disk (GB) & 4096 & 10 & 100 & 6400 & 80 \\
 & GPU & Yes & Yes & Yes & N/A & not provisioned \\
 & GPU memory (GB) & unknown & 4 & 8 & N/A & not provisioned \\
\midrule
\multirow{5}{*}{\texttt{bootstrap}} & CPU & unknown & 2 & 8 & 48 & 2 \\
 & RAM (GB) & unknown & 4 & 16 & 384 & 4 \\
 & Disk (GB) & unknown & 10 & 50 & 6400 & 40 \\
 & GPU & No & No & Yes & N/A & not provisioned \\
 & GPU memory (GB) & unknown & 0 & 8 & N/A & not provisioned \\
\midrule
\multirow{5}{*}{\texttt{pdf-to-podcast}} & CPU & 8 & 4 & 8 & 48 & 2 \\
 & RAM (GB) & 64 & 16 & 32 & 384 & 4 \\
 & Disk (GB) & 100 & 20 & 50 & 6400 & 40 \\
 & GPU & Yes & Yes & Yes & N/A & not provisioned \\
 & GPU memory (GB) & unknown & 8 & 24 & N/A & not provisioned \\
\bottomrule
\end{tabular}%
}

\vspace{0.5ex}
\footnotesize
\noindent\textit{Notes.} \textit{Specified} denotes the hardware requirements inferred by EnvAgent+ through repository analysis. \textit{Unknown} indicates that no reliable value could be determined from the repository analysis. \textit{Codex} and \textit{Claude} report hardware requirements inferred independently by the baseline agents and do not perform resource discovery or provisioning. \textit{Bare metal (TACC)} and \textit{KVM actual} report the hardware reserved by EnvAgent+, which is selected based on inferred requirements and available platform resources. Because TACC and KVM allocate fixed hardware configurations, the reserved resources may exceed the specified requirements. \textit{N/A} indicates that the resource is not applicable, whereas \textit{Not provisioned} indicates that no compatible resource was provisioned during the evaluation. These cases do not affect the aggregate provisioning results reported in Section~III-C.

\end{table}

\newpage
\section*{Data Availability}
The \MainAgent framework, \EnvBench evaluation suite, and our preprocessed
dataset compilation are open source and publicly available at
\href{https://github.com/EaminC/EnvAgent}{\faGithub\,
\texttt{github.com/EaminC/EnvAgent}}. Our experiments use publicly available
artifacts from the
\href{https://blog.chameleoncloud.org/posts/importing-github-repositories-to-trovi-a-step-by-step-guide/}{Chameleon Trovi Reproducible Research Collection}
and the
\href{https://huggingface.co/datasets/ByteDance-Seed/Multi-SWE-bench}{Multi-SWE-bench Cross-Language Collection}.

\end{document}